\documentclass{article}
\usepackage{authblk}
\title{IA generativa aplicada a la detección del cáncer a través de Resonancia Magnética}
\author[1]{Virginia del Campo}
\author[2]{Iker Malaina}
\affil[1]{Hospital de Urduliz, OSI Uribe, Osakidetza\\
Goieta 32, 48610 Urduliz, Bizkaia, Spain}
\affil[2]{Departamento de Matemáticas, FCT, UPV/EHU\\
Sarriena sn, 48940 Leioa, Bizkaia, Spain}

\usepackage{adjustbox}
\usepackage{pgfplots}
\pgfplotsset{compat=1.17}
\usepackage{amssymb}
\usepackage{amsmath}
\usepackage{graphicx}
\usepackage[misc]{ifsym}
\usepackage{comment} 
\usepackage{url}
\usepackage{tikz}

\urldef{\mailsa}\path|iker.malaina@ehu.eus|

\begin{document}

%\mainmatter  % start of an individual contribution

% first the title is needed

\begin{comment}
Comments: 9 pages, in Spanish
\end{comment}

\maketitle

\renewcommand{\abstractname}{English version of abstract}
\begin{abstract}
Cognitive delegation to artificial intelligence (AI) systems is transforming scientific research by enabling the automation of analytical processes and the discovery of new patterns in large datasets. This study examines the ability of AI to complement and expand knowledge in the analysis of breast cancer using dynamic contrast-enhanced magnetic resonance imaging (DCE-MRI). Building on a previous study, we assess the extent to which AI can generate novel approaches and successfully solve them. For this purpose, AI models, specifically ChatGPT-4o, were used for data preprocessing, hypothesis generation, and the application of clustering techniques, predictive modeling, and correlation network analysis. The results obtained were compared with manually computed outcomes, revealing limitations in process transparency and the accuracy of certain calculations. However, as AI reduces errors and improves reasoning capabilities, an important question arises regarding the future of scientific research: could automation replace the human role in science? This study seeks to open the debate on the methodological and ethical implications of a science dominated by artificial intelligence.

\end{abstract}
 -----
\renewcommand{\abstractname}{Other language version of abstract}
\begin{abstract}
La delegación cognitiva a sistemas de inteligencia artificial (IA) está transformando la investigación científica, permitiendo la automatización de procesos analíticos y el descubrimiento de nuevos patrones en grandes volúmenes de datos. Este estudio examina la capacidad de la IA para complementar y expandir el conocimiento en el análisis de cáncer de mama mediante resonancia magnética dinámica con contraste (DCE-MRI). Se parte de un trabajo previo, evaluando hasta qué punto la IA puede generar enfoques novedosos y resolverlos con éxito. Para ello, se utilizaron modelos de IA, específicamente ChatGPT-4o, para el preprocesamiento de datos, la generación de hipótesis y la aplicación de técnicas de clustering, modelado predictivo y análisis de redes de correlación. Se compararon sus resultados con los obtenidos manualmente, evidenciando limitaciones en la transparencia del proceso y en la precisión de ciertos cálculos. Sin embargo, a medida que la IA reduzca sus errores y aumente su capacidad de razonamiento, se plantea una reflexión sobre el futuro de la investigación científica: ¿podría la automatización reemplazar el papel humano en la ciencia? Este trabajo busca abrir el debate sobre las implicaciones metodológicas y éticas de una ciencia dominada por la inteligencia artificial.

\end{abstract}

\section{Introducción}

La delegación cognitiva se refiere al proceso mediante el cual los seres humanos transfieren tareas mentales o de procesamiento de información a herramientas o sistemas externos, como dispositivos tecnológicos o algoritmos de inteligencia artificial (IA). Esta práctica permite a las personas optimizar recursos cognitivos y enfocarse en actividades más complejas o creativas.

En el ámbito de la investigación científica, la IA ha emergido como una herramienta poderosa que facilita y potencia diversas etapas del proceso investigativo. Por ejemplo, existen aplicaciones como Rayyan, que utiliza IA y aprendizaje automático para agilizar procesos sistemáticos; Consensus, un motor de búsqueda con IA para encontrar información en artículos de investigación; y Research Rabbit , que ayuda en el descubrimiento de artículos similares. Estas herramientas no solo aceleran la recopilación y análisis de datos, sino que también permiten identificar patrones complejos que podrían pasar desapercibidos mediante métodos tradicionales .

Sin embargo, la delegación cognitiva a sistemas de IA conlleva riesgos que deben ser considerados. Uno de los principales es la posibilidad de que la IA genere respuestas incorrectas o sesgadas. Por ejemplo, se ha reportado que modelos de lenguaje como ChatGPT pueden tener porcentajes de error de hasta el 60\% en tareas complejas . Además, la dependencia excesiva de la IA puede inhibir el pensamiento crítico y la creatividad de los investigadores, ya que podrían aceptar sin cuestionar los resultados proporcionados por estas herramientas. Otro riesgo es la propagación de información incompleta o incoherente, lo que podría afectar la calidad y fiabilidad de las investigaciones. Por último, la falta de transparencia en los algoritmos de IA puede dificultar la comprensión de cómo se obtienen ciertos resultados, lo que plantea desafíos éticos y metodológicos en la investigación científica.

Para valorar las capacidades de la IA a día de hoy y hasta dónde la delegación congnitiva puede llevarnos al fin de la cienca como la conocemos, la idea del presente estudio es la de, partiendo de un artículo ya publicado \cite{iwbbio1}, utilizar la IA para ver hasta dónde es capaz de generar nuevos enfoques interesantes, y en caso de serlo, si puede resolverlos con éxito, esto es, medir sus capacidades como investigador. En resumen, el contexto del problema original es el siguiente: 

El cáncer de mama es una patología caracterizada por el crecimiento descontrolado de células en los tejidos mamarios, afectando principalmente a mujeres, aunque también puede presentarse en hombres. Su origen es multifactorial, incluyendo factores genéticos, hormonales y ambientales, como mutaciones en los genes BRCA1 y BRCA2 o la exposición prolongada a hormonas femeninas. En España, es el cáncer más común entre mujeres, con una incidencia estimada de 35,001 nuevos casos en 2023. A pesar de los avances en detección y tratamiento, sigue siendo la principal causa de muerte por cáncer en mujeres, con 6,528 fallecimientos previstos en 2023. Además, la recurrencia tumoral ocurre en hasta el 15\% de los casos, y en algunos estudios se ha observado progresión a metástasis en hasta el 21\% de los pacientes. La detección temprana es clave para mejorar el pronóstico del cáncer de mama, utilizando estrategias como mamografías, autoexploraciones y chequeos médicos regulares. Entre las técnicas más avanzadas, la resonancia magnética (MRI) permite obtener imágenes detalladas de los tejidos blandos mediante campos magnéticos y ondas de radio. En particular, la modalidad \textit{DCE-MRI} (Dynamic Contrast-Enhanced MRI) emplea un agente de contraste para evaluar la vascularización del tumor, proporcionando información crucial sobre su malignidad, planificación del tratamiento y respuesta a la terapia.

Basándonos en dicho problema y en la investigación llevada a cabo en \cite{iwbbio1} usando esos mismos recursos, el objetivo fundamental de este trabajo es explorar las capacidades para complementar los hallazgos mencionados a través del uso de IA, examinando su capacidad para identificar patrones, generar nuevos análisis y valorar la exactitud de los resultados obtenidos. Este enfoque permitirá no solo validar la capacidad de la IA para asistir en estudios científicos, sino también analizar sus limitaciones y posibles sesgos. De esta manera, se podrá establecer un marco más sólido para el uso de la delegación cognitiva en el análisis de datos biomédicos, asegurando que su aplicación contribuya a mejorar la calidad y fiabilidad de la investigación en cáncer de mama.

\section{Métodos}

\subsection{Resumen sobre la base de datos original}
En \cite{iwbbio1}, además de las imágenes de resonancia magnética (MRI) descritos más adelante, se recopilaron diversas variables de los pacientes. Se incluyeron datos demográficos, características tumorales, hallazgos en MRI, información sobre cirugía, radioterapia, respuesta tumoral, recurrencia, seguimiento y datos terapéuticos. Estas variables fueron extraídas de fuentes como registros médicos electrónicos, informes de biopsia, reportes radiológicos y notas clínicas. Entre los datos demográficos se registraron la edad al diagnóstico, estado menopáusico, raza/etnia y presencia de metástasis. Las características tumorales abarcaron información sobre receptores hormonales, subtipo molecular, puntuación Oncotype, estadificación TNM, grado tumoral y tipo histológico. Los hallazgos en MRI incluyeron multicentricidad, linfadenopatía e implicación de piel/pezón. También se documentaron detalles de cirugía y radioterapia, así como respuesta clínica y patológica al tratamiento.

La base de datos que se utilizó, está basada en la información obtenida a partir de imágenes de resonancia magnética (MRI) en cortes transversales de tejido mamario, adquiridas con resonadores de 1.5 o 3 Teslas en posición prona. Se emplearon secuencias MRI en formato DICOM, incluyendo: (i) secuencia T1 ponderada sin saturación de grasa, útil para caracterizar lesiones pero susceptible a artefactos (en la Figura 1 se ilustra un ejemplo de este tipo de imágenes en la detección de un cáncer de mama); (ii) secuencia T1 ponderada con saturación de grasa previa a la administración de contraste, que suprime la señal de grasa para resaltar diferencias sutiles en los tejidos; y (iii) entre tres y cuatro secuencias posteriores al contraste para evaluar la captación y dinámica tisular (en la Figura 2 se ilustra un ejemplo de este tipo de imágenes).

\begin{figure}
    \centering
    \includegraphics[width=0.6\textwidth]{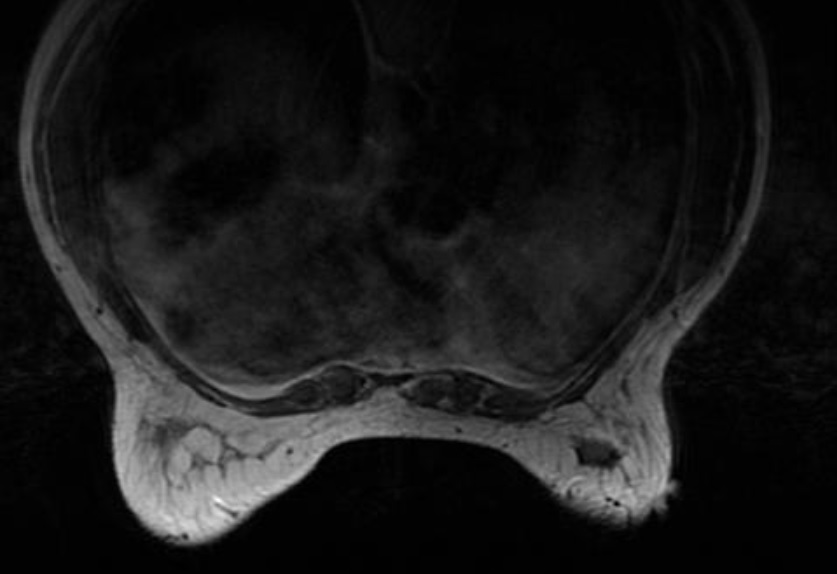}
    \caption{MRI axial T1 de una paciente con cáncer de mama (TR: 600 ms, TE: 10.08 ms; Frecuencia de imagen: 127.725 MHz).}
    \label{fig:axial}
\end{figure}

A partir de estas imágenes, se estableció un conjunto de características predictivas clasificadas según su origen y método de procesamiento. En \cite{iwbbio1}, nos basamos en estas características para evaluar la capacidad de la resonancia en la detección, considerando únicamente aquellas relevantes para la caracterización de subtipos de cáncer, y en particular, la recurrencia tumoral.

\begin{figure}
    \centering
    \includegraphics[width=0.6\textwidth]{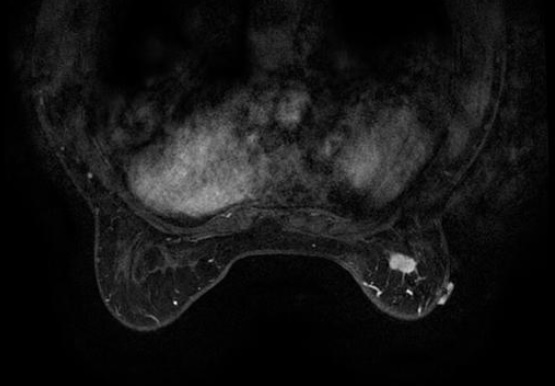}
    \caption{MRI dinámica axial post-contraste de una paciente con cáncer de mama (TR: 6,422 ms, TE: 2,644 ms; Frecuencia de imagen: 127,725503 MHz).}
    \label{fig:axial}
\end{figure}

\subsection{Preprocesamiento y Preparación de Datos}

En este trabajo, vamos a utilizar la versión GTP-4o del conocido ChatGPT. Su primera propuesta antes de sugerir análisis nuevos fue la de llevar a cabo una limpieza y preprocesamiento de la base de datos (ésta se hizo de manera manual en \cite{iwbbio1}). En particular, los pasos indicados fueron los siguientes:

\subsubsection{Limpieza y Normalización}
Para asegurar la integridad del conjunto de datos se llevaron a cabo los siguientes pasos:
\begin{enumerate}
    \item \textbf{Revisión de valores faltantes:}  
    Se inspeccionó la base de datos para identificar registros con valores incompletos o inconsistentes. En aquellos casos en que se detectaron datos faltantes, se eliminaron dichas variables del conjunto.
    
    \item \textbf{Detección de valores atípicos:}  
    Se analizaron las distribuciones individuales de las variables mediante diagramas de caja para identificar outliers. En la Figura 1 se ilustra, para dos de las variables de estudio, esta diferencia de escalas a través de los boxplot. Los valores atípicos fueron tratados mediante transformaciones logarítmicas para minimizar su impacto en los análisis posteriores.

    \item \textbf{Normalización y estandarización:}  
    Debido a la diversidad en las escalas de medición de las 529 variables, se procedió a la normalización (técnica Min-Max) de cada variable, con el objetivo de garantizar la comparabilidad entre las diferentes medidas.
\end{enumerate}

\subsubsection{Selección Inicial de Variables}
Previo a la aplicación de métodos de reducción de dimensionalidad y modelado predictivo, se efectuó una selección preliminar de variables, siguiendo estos criterios:
\begin{enumerate}
    \item \textbf{Eliminación de variables redundantes:}  
    Se calculó la varianza de cada variable y se descartaron aquellas con muy baja variabilidad, las cuales se consideraron poco informativas para la detección de patrones relevantes.
    
    \item \textbf{Detección de correlaciones fuertes:}  
    Se estimó la matriz de correlación para identificar grupos de variables altamente correlacionadas. En presencia de correlaciones elevadas, se optó por conservar únicamente una variable representativa de cada grupo, reduciendo la redundancia sin perder información sustancial.
    
\end{enumerate}

Esta fase de preprocesamiento sirvió para garantizar la calidad del análisis, permitiendo que las técnicas subsecuentes de reducción de dimensionalidad, clustering y modelado predictivo pudieran extraer información relevante sin verse afectadas por inconsistencias o sesgos en los datos.

\section{Resultados}
Una vez llevada a cabo la limpieza de la base de datos, y antes de valorar la capacidad de innovación y a su vez comprensión de ChatGPT, al comienzo de la sección de resultados se presenta un pequeño resumen de las técnicas ya implementadas en el trabajo que se utilizó como punto de partida para ChatGPT. El objetivo de esta parte es la de resaltar cómo . 

\subsection{Resumen de los análisis llevados a cabo en el trabajo previo \cite{iwbbio1}}
Para investigar la recurrencia y metástasis del cáncer de mama, primero se calcularon los Odds Ratios para evaluar la probabilidad de recurrencia y metástasis en cada subtipo (Luminal A, Luminal B, HER2 y TNBC).
A continuación, se evaluó la distribución de las variables mediante la prueba de Kolmogorov-Smirnov y, según su normalidad, se aplicaron ANOVA o Kruskal-Wallis. Se realizaron tres análisis: (1) identificación de variables distintivas entre subtipos, (2) detección de variables asociadas a recurrencia y (3) detección de variables asociadas a metástasis.

\subsection{Propuestas de investigación de la IA generativa}
Una vez alimentamos a ChatGPT con la base de datos y el trabajo previo, le solicitamos que analizase los métodos utilizados y que plantease herramientas o enfoques alternativos que dieran lugar a resultados novedosos y que fueran de interés. A continuación se muestra una lista de las propuestas que realizó:

\begin{enumerate}
    \item Métodos de Clustering
\begin{itemize}
    \item \textbf{K-means}
    \begin{itemize}
        \item Aplicar K-means sobre los datos o PCA reducido.
        \item Determinar \( k \) óptimo con método del codo o silueta.
    \end{itemize}
    \item \textbf{Clustering Jerárquico}
    \begin{itemize}
        \item Aplicar métodos aglomerativos (\textit{Ward}, enlace promedio).
        \item Identificar cortes óptimos en la jerarquía.
    \end{itemize}
\end{itemize}

    \item Modelos Predictivos Supervisados
\begin{itemize}
    \item \textbf{División de Datos y Validación Cruzada}
    \begin{itemize}
        \item Dividir datos (70\% entrenamiento, 30\% test).
        \item Validación cruzada \( k \)-fold.
    \end{itemize}
    \item \textbf{Algoritmos de Clasificación}
    \begin{itemize}
        \item Regresión Logística (interpretabilidad).
        \item Random Forest (relaciones no lineales, importancia de variables).
        \item SVM (alta dimensionalidad, kernels).
        \item XGBoost (patrones complejos, regularización).
        \item Entrenar modelos y optimizar hiperparámetros.
        \item Evaluar precisión, AUC-ROC, recall.
    \end{itemize}
    \item \textbf{Interpretabilidad del Modelo}
    \begin{itemize}
        \item Utilizar SHAP para analizar la contribución de cada variable.
    \end{itemize}
\end{itemize}

    \item Análisis de Redes de Correlación y Co-Expresión
\begin{itemize}
    \item \textbf{Matriz de Correlación}
    \begin{itemize}
        \item Calcular matriz de correlaciones (Pearson, Spearman).
        \item Visualizar con mapa de calor.
    \end{itemize}
    \item \textbf{Análisis de Redes y Detección de Módulos}
    \begin{itemize}
        \item Convertir matriz en red.
        \item Aplicar algoritmos de clustering (Louvain, Infomap).
    \end{itemize}
\end{itemize}

     \item Modelos de Supervivencia
\begin{itemize}
    \item \textbf{Análisis de Supervivencia con el Modelo de Cox}
    \begin{itemize}
        \item Aplicar el modelo de Cox.
        \item Calcular \textit{hazard ratios} e intervalos de confianza.
    \end{itemize}
\end{itemize}

\end{enumerate}

\section{Discusión}

Este trabajo representa una versión preliminar del estudio, en la cual se han empleado herramientas de inteligencia artificial para realizar ciertos cálculos. En la versión completa, se incluirán comparaciones entre los resultados obtenidos con ChatGPT y aquellos calculados manualmente, evidenciando las limitaciones actuales de la IA en este contexto específico. Los errores de cálculo y la falta de transparencia en algunos de los procesos internos de estas herramientas hacen que, por el momento, no sean completamente adecuadas para estudios científicos rigurosos sin supervisión humana.

Sin embargo, esta limitación es, probablemente, solo temporal. La evolución de la inteligencia artificial en los próximos años plantea un escenario en el que estos errores sean corregidos, lo que permitiría a las IA realizar análisis complejos con una precisión incluso superior a la humana. Este posible futuro nos obliga a reflexionar sobre el papel de la inteligencia humana en la ciencia.

Si la investigación científica llegara a estar completamente automatizada, con algoritmos capaces de formular hipótesis, diseñar experimentos y extraer conclusiones de manera autónoma, nos enfrentaríamos a un cambio paradigmático. ¿Sería esto un avance positivo o supondría una pérdida del valor inherente a la creatividad y el juicio crítico humano? Es innegable que la IA puede acelerar descubrimientos y optimizar recursos, pero surge la pregunta de si la ciencia debe ser únicamente una búsqueda de resultados o si su esencia radica también en el proceso intelectual detrás de ellos.

Además, existe el riesgo de que la automatización total de la investigación genere dependencia tecnológica y una falta de comprensión profunda de los fenómenos estudiados. Si los humanos dejan de ser los actores principales en la producción del conocimiento, ¿seguirán siendo capaces de interpretarlo y aplicarlo correctamente? En este sentido, la transición hacia una ciencia dominada por la IA deberá ser acompañada de una profunda reflexión ética y metodológica para asegurar que el conocimiento generado continúe siendo accesible y útil para la humanidad.

\end{document}